# Brownian motion with time-dependent friction and single-particle dynamics in liquids

Kirit N. Lad[1,*], Margi K. Patel[2] and Arun Pratap[2]

[1] Computational Condensed Matter Physics Laboratory, Department of Physics, Sardar Patel University, Vallabh Vidya Nagar-388120, Gujarat, India

[2] Condensed Matter Physics Laboratory, Applied Physics Department, Faculty of Technology & Engineering, The M. S. University of Baroda, Vadodara-390001, Gujarat, India

## Abstract

A microscopic theory of molecular motion in classical monatomic liquids, proposed by Glass and Rice [Phys. Rev. **176**, 239 (1968)], is revisited and extended to incorporate the dynamic friction in the Brownian description of the atomic diffusion in a mean-time-dependent harmonic force field. A modified, non-Markovian Langevin equation is utilized to derive an equation of motion for the velocity autocorrelation function with time-dependent friction coefficient. Numerical solution of the equation gives an excellent account of the velocity autocorrelation function in LJ liquids, liquid alkali and transition metals over a broad range of density and temperature. Derivation of the equation of motion leads to a self-consistent expression for the time-dependence of friction coefficient. Our results demonstrate that the nature of time-dependence of the friction coefficient changes dramatically with the liquid density. At low and moderate densities, the dynamic friction decays exponentially whereas it increases exponentially at high liquid densities. Our findings provide an opportunity to have a new outlook of the Brownian description of atomic dynamics in liquids.

*Corresponding Author: knlad-phy@spuvvn.edu

## I. Introduction

Theories for the description of the atomic motion in liquids are invariably linked directly or indirectly to the Brownian motion. The earliest theories of Brownian motion are credited to Einstein (1905) [1] and Smoluchowski (1906) [2] who established a connection between equilibrium fluctuations and viscous friction to obtain the diffusion coefficient of a large Brownian particle in a fluid. However, it was Langevin's classical theory [3] that provided a mathematically coherent framework encompassing the effect of thermal fluctuations on macroscopic dynamics. A detailed statistical mechanical treatment of Brownian motion and the Langevin theory can be found in a nice pedagogic review by Chandrasekar.[4] The essential concepts of the Langevin approach, though quite well-known, are necessary to be highlighted here in the context of the problem to be addressed in the present work. In the Langevin equation of motion for a massive Brownian particle, the force acting on it is split into two parts; the first being the systematic viscous drag due to the molecular friction and, the second rapidly fluctuating force due to collisions of the atoms of the fluids with the particle. It is assumed that the fluctuating force is completely random(white) and, it is not correlated to the instantaneous velocity of the particle. Both of these assumptions do not necessarily remain valid for the Brownian particles with the mass same as that of the surrounding fluid atoms because the timescale for the motion of the Brownian particle becomes similar to the characteristic timescales ($\sim 10^{-13}$–$10^{-12}$ s) of atomic dynamics in the fluid. In such cases, non-random time-dependent intermolecular forces do exist in the fluids. [5,6] Moreover, the viscous drag relies on an empirical friction constant which is usually obtained using the Stokes law suggesting overdamped motion at long times. In reality, friction is generally a time-dependent quantity owing to its origin from conservative intermolecular forces in the fluid and, understanding its origin still remains one of the open challenges of the physics of fluids.[7]

Kirkwood's generalized Brownian theory for liquids and solution was the first systematic approach that gave an explicit relationship between the friction constant and the intermolecular forces.[8] Mori [9] and Kubo [10] derived a generalized Langevin equation by taking into account the non-randomness in the force acting on the Brownian particle and the retarded effect of the frictional forces in terms of a time-dependent friction coefficient. As atomic motion in liquids closely conformed to the vibratory motion interspersed with diffusive motion revealed by the slow-neutron scattering experiments, several phenomenological models were proposed by considering the Brownian motion of the atoms in a non-random force field. Most notable among these models are the simple harmonic well (SHW) model [11], the itinerant oscillator model [12,13], a stochastic model [14], hindered-translator model [15]. These models are based on an underlying assumption of solid-like behaviour of the liquids at very short timescale and it also assume static non-random force field. A disordered solid-like picture of a liquid was also the basis of Zwanzig's idea of describing the atomic motion as oscillations about a local minimum on the potential energy surface and characterizing the self-diffusion process as infrequent jumps between different local minima.[16] This approach, which differs conceptually from the kinetic theory, has been reformulated in many different ways. The most famous of these, the instantaneous normal modes (INM) approach [17-19], has been found to be very useful for analysing the short-time dynamics of liquid and has also been extended to study the self-diffusion. Despite its ability to provide quantitatively accurate results for self-diffusion coefficients in a variety of liquids, it has been often criticized due to some inherent ambiguities related to the self-diffusion process in the configuration space. [20,21] To circumvent such issues in the description of the short-time dynamics as well as self-diffusion in liquids, hybrid theories and models [22-25] have been formulated by combining the kinetic theory and the INM approach. Most of these models and theories, involving assumption of quasi-phonon excitation, are too complex in its formulation and provide an elusive physical picture of microscopic motion in liquids.

A more realistic description of the molecular motion in liquids has been given by the microscopic theory proposed by Glass and Rice (GR).[26] It is devoid of the assumption of the quasi-phonon excitation and limitations of the kinetic theory. It addresses the non-random time-dependent forces in the liquid by considering a Brownian particle diffusing in a mean time-dependent force field and introducing an additional systematic force term to the Langevin equation. (see Sec. II) The idea is based on the work of Rice and co-workers [27] where two forces are identified as: (i) a rapidly fluctuating force arising from the molecular motion in the soft long-range part of the intermolecular potential and, (ii) the short-range strongly repulsive forces during core collisions leading to the "backscattering" of the molecules. The core collisions are assumed to be dynamically uncorrelated. While the assumption about the rapidly fluctuating forces is retained in the GR theory, it considers the strongly repulsive core collisions to be dynamically correlated and its effects are represented by a time-dependent average force field. Decay of this mean force is associated with the molecular relaxation time. The modified Langevin equation has been utilized to derive an equation of motion in terms of the velocity autocorrelation function (VAF). It involves two rate parameters related to two different characteristic timescales. One corresponds to the molecular relaxation rate, $\alpha$, which corresponds to the time for which an atom remains in the same local environment (nearest-neighbour cage). The associated time is different from the Maxwell relaxation time (τ) and similar to Frenkel's interpretation as the average time for which the molecule remains attached to the same equilibrium position.[28] It is equivalent to the time for which a local relaxation event involving large-scale rearrangement of cage atoms last and, it is of the order of the Debye vibration period $τ_0 ≈ 0.1$ ps. [29] It has also been interpreted as an attempt time to undergo relaxation events over some typical length scale. [30] Although $τ ≈ τ_0$ at high temperatures, $τ ≈ 10^3$ s at the glass transition. [29] The other timescale is related to the friction constant, β, which represents the damping of the oscillatory motion of the atoms inside the nearest-neighbour cage on account of "backscattering" from the wall of the cage. To obtain an analytical solution of the equation of motion, i.e., VAF, GR assumed α and β to be equal. It stems from a physical point of view that the decay of the mean force results from the loss of correlation among the surrounding atoms of the first coordination shell due to diffusion. This assumption is quite reasonable for a monatomic, homogeneous liquid at low density where the relaxation times for all the atoms are same. GR model gives VAF for liquid argon that agrees well with the computer simulation result [31] up to the backscattering region. But the subsequent oscillations are absent. It primarily implies that the assumption of α being equal to β misses some important physics of atomic dynamics. This point has been emphasized in the studies of the atomic dynamics in liquid metals and alloys where a generalized GR solution is provided subject to the more realistic condition $α ≠ β$. [32] The reported VAF results indicate that the generalized GR solution also fails to give a good quantitative account of the backscattering region and the subsequent oscillations.

In the present study, we revisit the GR theory to find any shortcomings in its formulation which, if addressed adequately, would provide better quantitative results and hence, a better description of the molecular motion in liquids. Apparently, the first shortcoming is the assumption of the friction coefficient (β) to be independent of time. It is questionable when the mass of a Brownian particle is similar to the surrounding particles. It also misses to capture the possible effects of the dynamic correlations on molecular friction. In fact, GR themselves suggested the possible extension of their approach to include the dynamical friction for the case where the period of oscillation of the long-range, rapidly fluctuating soft forces is not sufficiently short.[26] We show that the incorporation of time-dependent friction in GR theory leads to an equation of motion in terms of VAF with its solution encompassing three possible physical scenarios for the time-dependence of molecular friction. The remaining paper is organized as follows. The essential concepts of the GR theory, which is necessary for the present work, are summarized in Sec. II. Sec. III deals with the derivation of the modified equation of motion in terms of VAF incorporating time-dependent friction. The results of VAFs for the LJ systems, liquid alkali metals and transition metals are presented in Sec. IV and are further discussed in detail in Sec. V. Major highlights of the present work are summarized as conclusions in Sec. VI. Details of molecular dynamics simulations performed for the liquid transition metals are given in the Appendix.

## II. GLASS AND RICE FORMULATION OF MODIFIED LANGEVIN EQUATION

The modified Langevin equation obtained by GR [26] is

$$m\frac{d\vec{v}}{dt} = -m\beta\vec{v}(t) + \vec{F}(\vec{R},t) + \vec{A}(t) \tag{1}$$

where $\vec{v}$ is the instantaneous velocity of the Brownian particle of mass $m$. The first term on the right-hand side of Eq. (1) is the systematic force term due to viscous drag where $\beta$ is the friction coefficient. The second term is another systematic force term representing the time-dependent average force field due to the effects of the dynamically correlated, strongly repulsive core collisions. $\vec{A}(t)$ is the stochastic force that is assumed to result from rapid fluctuations in the soft part of the intermolecular potential field. While a detailed theoretical formulation of $\vec{F}(\vec{R},t)$ is given by GR [26], a brief summary of important points about this non-random force is in order at this place.

First of all, the mean force, $\vec{F}(\vec{R},t)$, is defined in terms of an ensemble-averaged, two-particle, conditional distribution function, $P(\vec{R}',t|\vec{R},t)$, giving the conditional probability density that a particle is at the point $\vec{R}'$ at time $t$, $(\vec{R}',t)$, given that a different particle which was at $(0,0)$ has moved to $(\vec{R},t)$. In the absence of external forces and assuming pairwise decomposable interactions in liquids, $\vec{F}(\vec{R},t)$ has been given to be

$$\vec{F}(\vec{R},t) = -\int d\vec{R}'\vec{\nabla}V(\vec{R}'-\vec{R})\,P(\vec{R}',t|\vec{R},t) \tag{2}$$

where $V(\vec{R}'-\vec{R})$ is the two-particle potential energy.

To solve Eq. (2), one needs to obtain the direct solution of the Bogoliubov-Born-Green-Kirkwood-Yvon (BBGKY) hierarchy, which itself is an ordeal. GR circumvented this problem by employing an alternative approach of relaxation methods for the representation of particle motion in liquid. This approach does not specify the details of dynamical forces which act on a subsystem of particles. However, it is assumed that the forces present in the liquid tend to drive the system towards its equilibrium configuration. So, without considering the exact nature of the propagation and destruction of correlations in the liquid, GR proposed a simple relaxation approximation for $P(\vec{R}',t|\vec{R},t)$ as

$$\frac{d}{dt} P(\vec{R}',t|\vec{R},t) = \alpha[\rho g(\vec{R}' - \vec{R}) - P(\vec{R}',t|\vec{R},t)] \tag{3}$$

where, $\alpha$ is a parameter corresponding to the molecular relaxation rate. An approximate expression for $P(\vec{R}',t|\vec{R},t)$ obtained by solving Eq. (3) is

$$P(\vec{R}',t|\vec{R},t) = e^{-\alpha t}\rho g(\vec{R}') + (1 - e^{-\alpha t})\rho g(\vec{R}' - \vec{R}) \tag{4}$$

On substituting Eq. (4) in Eq. (2) leads to

$$\vec{F}(\vec{R},t) = e^{-\alpha t} \int d\vec{R}' \rho g(\vec{R}') \vec{\nabla} V[\vec{R}'(t) - \vec{R}(t))] \tag{5}$$

As the forces being considered are short-ranged, $\vec{\nabla} V$ will be non-zero only for small $\vec{R}'(t) - \vec{R}(t)$. Also, $\vec{R}(t)$ is small relative to the intermolecular spacing at the diffusion timescale. For small $\vec{R}(t)$, Eq. (5) has been expanded to yield [26]

$$\vec{F}(\vec{R},t) = -\vec{R}(t) e^{-\alpha t} \frac{\langle \nabla^2 V(R) \rangle}{3} + O(R^2) \tag{6}$$

where

$$\langle \nabla^2 V(R) \rangle = 4\pi \rho \int_0^\infty dR R^2 g(R) \left( \frac{\partial^2 V}{\partial R^2} + \frac{2}{R} \frac{\partial V}{\partial R} \right) \tag{7}$$

It should be noted that Eq. (6) differs slightly from that given by GR (Eq. 22 in Ref. 26) where the factor (1/3) does not appear. Also, in the equation given by GR, $\rho$ appears as a multiplication factor in the first term of Eq. (6) which makes the equation dimensionally incorrect. So, considering it as a possible typographical mistake, we have removed it.

Using the relation [33]

$$\omega_0^2 = \frac{\langle \nabla^2 V(R) \rangle}{3m} \tag{8}$$

Eq. (6) can be written as

$$\vec{F}(\vec{R},t) = -m\omega_0^2 \vec{R}(t) e^{-\alpha t} + O(R^2) \tag{9}$$

Eq. (9) describes a spherically symmetric, harmonic force field that decays exponentially with time. For the case of $\alpha = 0$, it turns out to be static-harmonic force field i.e., the SHW model. [11] In the present work, we use Eq. (9) to derive the equation of motion in terms of VAF as described in the following section.

## III. VELOCITY AUTOCORRELATION FUNCTION AND EQUATION OF MOTION

The normalized velocity autocorrelation is defined as [26]

$$\psi(t) = \frac{\langle \vec{v}(0) \cdot \vec{v}(t) \rangle}{\langle v^2(0) \rangle} \tag{10}$$

It is subjected to the following boundary conditions:

$$\lim_{t \to 0} \psi(t) = 1 \tag{11a}$$

$$\lim_{t \to 0} \frac{d\psi}{dt} = 0 \tag{11b}$$

$$\lim_{t \to 0} \frac{d^2\psi}{dt^2} = -\frac{\langle \nabla^2 V \rangle}{3m} = -\omega_0^2 \tag{11c}$$

where, $\omega_0$, is the liquid-characteristic frequency associated with the harmonic potential well. To obtain the equation of motion in terms of VAF, multiply Eq. (1) by $\vec{v}(0)/v^2(0)$ and take the ensemble average to get

$$\frac{d\psi}{dt} = -\beta\psi + \frac{\langle \vec{v}(0) \cdot \vec{F}(\vec{R},t) \rangle}{m \langle v^2(0) \rangle} \tag{12}$$

Substituting Eq. (9) for $\vec{F}(\vec{R},t)$, we get [34]

$$\frac{d\psi}{dt} = -\beta\psi - \omega_0^2 e^{-\alpha t} \int_0^t dt' \psi(t') \tag{13}$$

GR considers the friction coefficient to be a meaningful concept only for times which are longer than the period of the rapidly fluctuating forces. Hence, β is taken to be a time-independent parameter. In this case, Eq. (12) fails for the times $t \to 0$. This failure has been assumed to be restricted to only a small-time interval near $t = 0$ and Eq. (12) remains valid over a large interval of time. We believe that such reasoning for β to be independent of time is inadequate for a situation where a particle moves in a mean time-dependent force field and the strong, short-ranged repulsive collisions are considered to be dynamically correlated. In fact, we show in the following derivation how the assumption of time-independent β leads to an equation of motion that is inconsistent with the condition in Eq. (11c).

On differentiating Eq. (13) with respect to time and further simplification, one can arrive the equation given by

$$\frac{d^2\psi}{dt^2} + (\alpha + \beta)\frac{d\psi}{dt} + (\omega_0^2 e^{-\alpha t} + \alpha\beta)\psi = 0 \tag{14}$$

Eq. (14) represents the equation of motion in terms of VAF derived by GR. While GR assumes $\omega_0$ to be just a parameter and replaces it by $\omega$, we retain it for better physical perspective. For the limit $t \to 0$, Eq. (14) yields

$$\lim_{t \to 0} \frac{d^2\psi}{dt^2} = -(\omega_0^2 + \alpha\beta) \tag{15}$$

Thus, Eq. (14) does not satisfy the condition in Eq. (11c). As $\alpha$ is a non-zero parameter, the only possible way Eq. (14) can comply with this condition is to consider $\beta = 0$ at $t = 0$. It is clearly not the case in GR formalism where $\beta$ is assumed to be constant over the entire time interval of interest. We show in the following derivation how considering time-dependent friction coefficient, $\beta(t)$, resolves this inconsistency. It also allows us to get a self-consistent expression giving the time dependence of $\beta$.

Consider $\beta$ to be a time-dependent variable in Eq. (12). Differentiation of this equation with respect to time and further simplification gives

$$\frac{d^2\psi}{dt^2} + (\alpha + \beta(t))\frac{d\psi}{dt} + \left(\omega_0^2 e^{-\alpha t} + \frac{d\beta(t)}{dt} + \alpha\beta(t)\right)\psi = 0 \tag{16}$$

Employing the conditions in Eq. (11) for the limit $t \to 0$, Eq. (16) leads to

$$\lim_{t \to 0} \frac{d^2\psi}{dt^2} = \lim_{t \to 0} \left(\omega_0^2 e^{-\alpha t} + \frac{d\beta(t)}{dt} + \alpha\beta(t)\right) \tag{17}$$

It implies that Eq. (16) would satisfy Eq. (11c) if

$$\frac{d\beta(t)}{dt} + \alpha\beta(t) = 0 \tag{18}$$

Solving Eq. (18) and considering $\beta = \beta_0$ at $t = 0$, we get

$$\beta(t) = \beta_0 e^{-\alpha t} \tag{19}$$

Explanation for the choice of $\beta = \beta_0$ at the initial time $t = 0$ is necessary at this place. It should be noted that $\beta_0$ is the initial value of the friction coefficient at a time that corresponds to the onset of molecular friction. It has been shown in a recent study of onset of molecular friction in liquids [7] that, on all scales, friction emerges as a phenomenon that is non-local in time. It implies that the magnitude of friction cannot be inferred from the instantaneous behaviour of molecular trajectories. The same study also demonstrates that friction in liquids emerges abruptly at a characteristic frequency (being $\omega_0$ in the present study) for which viscous liquid appears as non-dissipative, elastic solid. Beyond the characteristic frequency the friction decays exponentially fast to zero. In this context, the following equation derived for the SHW model [11] would give physically adequate estimate of $\beta_0$.

$$\beta_0 = \left(\frac{mD}{k_B T}\right)\omega_0^2 \tag{20}$$

Now, using Eq. (19) in Eq. (16), we obtain the equation of motion in terms VAF as

$$\frac{d^2\psi}{dt^2} + (\alpha + \beta_0 e^{-\alpha t})\frac{d\psi}{dt} + \omega_0^2 e^{-\alpha t} \psi = 0 \tag{21}$$

Eq. (19) and (21) are the main features of the present work. Eq. (19) gives time-dependent friction coefficient and indicates connection between the short-time dynamics and the onset of the hydrodynamic regime. Eq. (21) is a more general equation of motion for the Brownian description of atomic motion in liquids.

It is evident that the knowledge of the parameters $\omega_0$, $\beta_0$ and $\alpha$ is essential for obtaining solution of Eq. (21). Although the physical meaning of these parameters has already been mentioned at relevant places in this section, some important remarks are necessary to highlight the physical connotations of these parameters in the context of time-dependent friction. In the usual time-independent friction case, $\omega_0$ is considered to be the Einstein frequency that can be derived using Eq. (11c). When the friction is dynamic, $\omega_0$ refers to the liquid-characteristic frequency which is different from the Einstein frequency. $\beta_0$ is the friction coefficient in the hydrodynamic limit and, it can be obtained using the Einstein relation, $\beta_0 = \frac{k_B T}{mD}$, if the diffusion coefficient (D) is known. However, the SHW model (Eq. 20) provides better initial estimate of $\beta_0$. In the dynamic friction case, $\beta_0$ corresponds to onset friction and it is expected to be different from that obtained using any of the ways suggested above. The molecular relaxation rate, $\alpha$, can be estimated utilizing the relation $\alpha\beta = \frac{\langle \nabla^2 V \rangle}{3m} = \omega_0^2$. [35] As $\omega_0$, $\beta_0$ and $\alpha$, obtained using above relations, do not correspond to the dynamic friction situation, we will consider it as fitting parameters. Nevertheless, the values of these parameters, given by the above relations, serve as physically meaningful initial guess for each. It should be noted that the molecular relaxation rate, $\alpha$, governs the mean-time dependent force field (Eq. 9) experienced by the Brownian particle as well as the time dependence of the friction force (Eq. 19). The two systematic force parts in the Langevin equation (Eq. 1) are no longer independent and, the interplay of the two plays an important role in the propagation and destruction of dynamical correlations in liquids. As the molecular relaxation rate depends on temperature and density of the liquid, the values of $\alpha$ would provide useful insight of the dynamical correlations and the diffusion process in the liquids.

For the simplest case of $\alpha = 0$, Eq. (21) turns out to be

$$\frac{d^2\psi}{dt^2} + \beta_0 \frac{d\psi}{dt} + \omega_0^2 \psi = 0 \qquad (22)$$

Eq. (22) is the equation of motion for a Brownian particle diffusing in a static harmonic well (SHW model) and has a well-established analytical solution giving the VAF.[26] It corresponds to Markovian diffusion process where the friction coefficient is independent of time. For $\alpha \neq 0$, it is not possible to obtain a trivial analytical solution of Eq. (21). However, it can be solved numerically to obtain VAF. In the present work, we have used a general numerical differential equation solver, NDSolve, a built-in language function in Wolfram Mathematica. The solutions for the GR model are also obtained using NDSolve. Numerical solution of Eq. (21) leads to two possibilities for the time dependence of the friction coefficient corresponding to values of $\alpha \neq 0$ (Eq. 19). For $\alpha > 0$, the friction coefficient would decay exponentially whereas it would increase exponentially for $\alpha < 0$. In the following section, we demonstrate in detail how $\alpha >$ 0 gives a better description of VAF in low density fluids whereas $\alpha < 0$ is inevitable to obtain consistent results of VAF for high density liquids.

## IV. RESULTS

Present theoretical framework for the Brownian description of the atomic motion in fluids encompasses Markovian and Non-Markovian Langevin equations involving time-independent and time-dependent friction, respectively. To expound the physical significance of different physical scenarios emerging on numerical solution of Eq. (21) and to demonstrate a broader scope of its applicability, we present results of VAF for various systems at different densities and temperatures: (i) Lennard-Jones (LJ) fluids, (ii) liquid alkali metals (Li, Na, K) and, (iii) liquid transition metals (Cu, Ni, Fe). As mentioned in the previous section, the input parameters $\omega_0$, $\beta_0$ and $\alpha$ have been considered to be fitting parameters so as to obtain VAF which is in the best agreement with the molecular dynamics results. However, we do obtain the initial estimates for $\omega_0$ and $\beta_0$ using Eq. (8) and (20), respectively for all the studied liquids except the LJ systems. We denote these parameters as $(\omega_0)_{fit}$ and $(\beta_0)_{fit}$. For LJ systems initial estimates for these parameters were obtained using the Einstein relation $\beta_0 = \frac{k_B T}{mD}$.

Our choice of LJ fluids has been motivated by the availability of MD results for these systems, reported by Bembenek and Szamel [36], at various fluid densities at a constant temperature (T = 1.5 in reduced unit). We employ present formulation to obtain VAFs that best fit the MD results as shown in Fig. 1. The parameters $\omega_0$, $\beta_0$ and $\alpha$ along with its optimized values are listed in Table I. VAFs obtained using the GR and SHW models have also been depicted to emphasize the implication of static friction used in these models vis-à-vis the use of dynamic friction in the present model for the Brownian description of single-particle dynamics in liquids. It can be observed that the present model with dynamic friction gives VAFs that are in the closest agreement with the MD results at the LJ fluid densities. This can be further validated from the residue plot ($\Delta\Psi = \Psi_{MD} - \Psi_{Model}$) in Fig. 1(d). However, a closer look at the residue plots for the present model results at different densities in Fig. 1(e) reveals that the agreement between the MD simulation and the present model results becomes poorer at lower densities. Prima facie, this may seem to be due the use of shifted LJ potential for MD in ref. 36 whose results are used here for comparison. As reported in ref. 36, the potential minimum V($r_m$) value is 0.9448 at $r_m$ = 1.1228σ with $r_{cut}$ = 2.5σ. V($r_m$) is 5% smaller than its usual value (1.0) for the standard LJ potential and $r_m$ remains almost unchanged compared to $2^{1/6}$σ ~ 1.1225σ. Toxvaerd and Dyre [37] have shown that a significant shifting of LJ potential due to setting up $r_{cut}$ from $2.5\sigma$ to $1.5\sigma$ have a very little effect on the dynamics of the system. Thus, the apparent lack of disagreement of the present model results with the MD data at lower densities could not be attributed to the use of the shifted LJ potential. The explanation for the disagreement lies in the assumption used in deriving Eq. (9) where the force $\vec{F}(\vec{R},t)$ is considered to be short-ranged and non-zero only for $\vec{R}(t)$ smaller than the average intermolecular spacing at the diffusion timescale. While this assumption is valid at higher densities and short times, it may not hold at lower densities and long times.

Our results provide very important insight of the change in the nature of the dynamic friction with the change in liquid density. For LJ liquids with densities 0.2 and 0.4, where the VAFs decays monotonically (and likely exponentially) (Fig. 1(a) & (b)), values of the parameter $\alpha$ along with Eq. (19) clearly indicate that the dynamic friction decreases exponentially. Also, the decrease in friction becomes faster with increase in liquid density. At the density 0.84, the VAF is similar to that of a typical high-density liquid exhibiting a negative backscattering region (Fig. 1(c)). For this case, we have found that the present model gives VAFs in agreement with the MD data only if $\alpha < 0$. The inset in Fig. 1(c) demonstrates that the VAFs obtained using the present are way off from the MD results for $\alpha > 0$. Negative $\alpha$ implies that the dynamic friction increases exponentially with time. The change in the nature of dynamic friction with the change in the liquid density is a significant finding insofar as it helps to unravel the change in velocity correlations and short-time dynamics. Deferring an elaborate discussion to the next section, we further consolidate present findings by investigating the VAFs in real high-density liquids such as liquid alkali metals and transition metals. In order to get the initial estimates of the parameters $\omega_0$ and $\beta_0$, for obtaining VAFs using present model, it is necessary to know $\langle \nabla^2 V \rangle$ (Eq. 7). To this end, we obtain the derivatives of the effective pair potentials in pseudopotential formalism [38] using evanescent-core pseudopotential [39] and Ichimaru-Utsumi screening function [40]. Fig. 2, 3 and 4 show VAFs for liquid Li, Na and K at different temperatures, respectively. The initial estimates for the parameters $\omega_0$, $\beta_0$ and its optimum values $(\omega_0)_{fit}$, $(\beta_0)_{fit}$, giving VAFs in good agreement with the MD results [41-43], are listed in Table. II.

For the liquid transition metals Cu, Ni and Fe, we have performed classical MD simulations at various temperatures. Necessary details of MD simulations are described in the Appendix. VAFs for Cu, Ni and Fe at different temperatures are presented in Fig. 5, 6 and 7 respectively. In this case, $\langle \nabla^2 V \rangle$ has been derived using the effective pair potentials obtained using the Wills-Harrison formulation [44] for the transition metals, employing Ashcroft pseudopotential [45] and Ichimaru-Utsumi screening function [40]. The parameters for all the studied transition metals are given in Table III. The key observations from the results of VAFs for the studied alkali and transition metals are as follows: (1) present model with dynamic friction, in general, gives an excellent account of the decay of velocity correlations in all the systems at all temperatures under investigation. However, at temperatures near or below the melting point, the oscillatory tail following the backscattering region is a little overestimated whereas at temperatures near the boiling point it is underestimated. This can be clearly understood from the VAFs for liquid Na at 900 K which is relatively closer to the boiling point (1155 K) (Fig. 3(a)) and 380 K (Fig. 3(b)) which is closer to its melting point (370 K). For liquid Ni, it can be observed from Fig. 6 that the deviations in the oscillatory tail becomes relatively pronounces near and below its melting point ($T_m$ = 1728 K). Similar observation can be made for liquid Fe from Fig. 7. (2) For all the studied liquid metals, values of the parameter $\alpha < 0$ (Table II & III) as observed in case of LJ system. (3) The Einstein frequency, $\omega_0$, is nearly equal to $(\omega_0)_{fit}$. Whereas the values of the friction coefficient $\beta_0$ are significantly less than $(\beta_0)_{fit}$. (Table II & III)

## V. DISCUSSION

Our results for the model LJ systems, liquid alkali and transition metals suggest that the present model (Eq. 21), based on the Langevin equation (Eq. 1) with time-dependent friction coefficient (Eq. 19) and mean time-dependent force (Eq. 9) gives a very good account of the short-time, single particle dynamics in fluids. The effect of the mean time-dependent force on the velocity correlations and the molecular friction is reflected in the VAF. It is evident from the results for the LJ systems that the nature of dynamic friction changes with the density of the fluid. At lower and moderate densities, the friction coefficient decreases exponentially and the VAF exhibits monotonic exponential decay. At these densities, the dynamical correlations are mainly governed by the binary collisions between the atoms where the average time during which two atoms interact (mean collision time) is much shorter than the average time between successive collisions (mean free time).[25] These two timescales become comparable in high density liquids where the effect of many-body correlations on the single-particle motion becomes significant and the binary collision picture fails to describe the short-time dynamics except for hard sphere liquids.[36] Values of the parameters $(\omega_0)_{fit}$ and $(\beta_0)_{fit}$ for the LJ systems (Table I) give a convincing testimony to these findings. These values will make more sense if we note that $\omega_0$ is interpreted as an effective frequency of atomic interactions (collisions) [46] and, $\beta_0$, the friction coefficient, is associated with the duration of interaction between the atoms during the collisions. In accordance with the above arguments, $(\omega_0)_{fit}$ is significantly smaller than $(\beta_0)_{fit}$ for the LJ fluids of density 0.2 and 0.4 whereas $(\omega_0)_{fit} \sim (\beta_0)_{fit}$ for ρ = 0.84.

The effect of dynamically correlated interactions of atoms on the single-particle diffusion in high density liquids manifests as two distinct features in the VAF. First is the negative region, also known as "backscattering" region, due to the caging of the diffusing atom by its nearest-neighbour atoms. The other is the long-time tail embodying the density waves excited by the initial push given to the medium by the moving atom. [47] The oscillatory or non-oscillatory long-time tail of VAF has been shown to be strongly connected to the softness or hardness of the repulsive core of the pair potential. [48-50] The pronounced oscillatory tail observed in the VAFs of liquid metals can be attributed to the existence of short-wavelength longitudinal modes of propagating collective excitations which are weakly damped due to the soft repulsive core. [47] Such short-wavelength density waves are strongly damped in the fluids with LJ-type hard repulsive core (like argon) and gives rise to the smoothly varying, non-oscillatory (exhibiting $t^{-3/2}$ dependence) tail of VAF. Unlike high density LJ liquid, the results of GR model for the liquid metals (Fig. 2-7) exhibit the most pronounced backscattering region and, in most cases, an unusually long and large amplitude oscillatory tail. This could be primarily due to the absence of the dynamic friction in the GR model. On the other hand, the presence of exponentially increasing dynamic friction in the present model leads to significant damping of the oscillations in the VAF that is consistent with the MD results.

Our, results for the LJ liquid (ρ = 0.84) and the liquid metals demonstrate that the present model would give a nearly complete account of the VAF, including the negative region and the long-time tail if the parameter $\alpha < 0$. This, according to Eq. (19), implies that the dynamic friction increases exponentially with time. It is significantly different from the usually reported time-dependence of the dynamic friction where the friction coefficient shows sudden increase from zero at $t = 0$ to a maximum value and subsequent exponentially decay to a constant hydrodynamic value. [49,50] For high density liquids with continuous interactions, presently observed time-dependence of the friction coefficient with slow increase from a non-zero value near $t \to 0$ to a gradual rapid increase is qualitatively more logical. The time-dependence of the dynamic friction, observed in the present case, is significant for atomic dynamics in the short-time, high-frequency elastic limit. In a recent study of the origin of the molecular friction in the liquids, Straube et al [7] have reported that the friction in liquids emerges abruptly at a characteristic frequency, beyond which viscous liquids appear as non-dissipative elastic solids. From the high frequency limit where the dynamic friction abruptly sets in, it increases exponentially till it slowly saturates to its hydrodynamic limit near $\omega_0$. It has been shown that the molecular friction depends on the complex interplay of fast and slow processes on very disparate timescales and; the driving mechanism behind the onset of the friction is not in the structural relaxation (often associated to cage relaxation). At very short-times and high frequencies, irrespective of caging of atoms occurs or not, the fast, yet irreversible momentum transfer to the neighbouring atoms drives the onset of the friction. [7] Thus, it is evident that the parameter, $\alpha$, corresponding to the molecular relaxation rate, governs the dynamic friction and the velocity correlations in liquids. While the physical scenarios for the short-time dynamics and the velocity correlations for the cases of $\alpha \geq 0$ are clear, the negative values of $\alpha$ for the high-density liquids seems to be counterintuitive. However, the fact that $\alpha < 0$ is inevitable to get consistent results of VAF in high density liquids makes a strong case for the interpretation of an underlying physical scenario. Due to non-availability of a trivial analytical solution of Eq. (21), it is difficult to draw any direct inferences about the implications of $\alpha$ on the short-time dynamics and relaxation mechanism. Nevertheless, the theoretical approaches, based on the instantaneous normal modes, guide us to seek a plausible explanation for $\alpha < 0$ as reminiscent to the existence of the imaginary eigenmodes in the short-time vibrational spectrum of liquids. A study of molecular origin of friction in liquids emphasizes that the short-time components of the friction arise from the microscopically well-defined INMs of the liquid. [53] According to a recent theoretical approach [54], an exact solution of the generalized Langevin equation for a normalized autocorrelation function ($C(t)$) of a classical many-body system can be expressed as an infinite sum of exponential (real and/or complex) functions as $C(t) = \sum_{j=1}^{\infty} I_j \exp(z_j t)$, where $I_j$ and $z_j$ are mode amplitudes and frequencies respectively, in an inherent eigenmode representation. These eigenmodes can be associated to relaxation channels in the system. As remarked by Bellissima et al [54], if $I_j$ and $z_j$ are complex quantities, the corresponding modes and its complex conjugate are both present in the series and, taken together, they represent an exponentially damped oscillation. On the other hand, real $I_j$ and $z_j$ define a purely exponential decay. In the context of these observations, our results of VAFs for high density liquids hint at the likelihood of the negative $\alpha$ to be among the arguments of the complex exponential functions corresponding to the imaginary eigenmodes. In an instantaneous normal mode analysis of liquid Na [55], the exponent functions derived from the INM density of states are found to fit well with a binomial frequency for the imaginary-frequency lobe and a three-term polynomial for the real-frequency mode. Also, the imaginary-frequency modes in the high-frequency end are found to be localized modes. Noting that the present formulation considers Brownian motion of an atom in a mean time-dependent harmonic force field (Eq. 9), we further extend our argument (although at the risk of oversimplification) that negative $\alpha$ is associated to the existence of imaginary INMs (unstable modes) in Zwanzig's model [16] dividing the liquid configuration space into "cells" - each associated with a local minimum on the potential energy surface. The liquid's configuration oscillates harmonically in one of these cells for a definite period of time and, on gaining enough kinetic energy cross a saddle point on the potential energy surface to make a transition of another cell with different local minimum.[19] The transitions over the saddles on the potential energy surface are characterized as the imaginary (unstable) eigenmodes in the density of states of INMs.[17] For $\alpha < 0$, Eq. (9) indicates increase in the mean-time-dependent harmonic force with time. In a constant density condition, it implies that the liquid explores deeper local minimum with time. It can be considered analogous to the saddle point transitions on the potential energy surface and hence, the imaginary INMs. On a cursory note, it can be observed that the value of $\alpha$ for liquid metals, in general, increases at lower temperatures and it might signify lower potential energy minima being explored by the liquids. To end this section, we would to point out that the GR theory, not explicitly considering "quasi-phonon" excitations, does acknowledge the possible utility of Zwanzig's picture for the adequate representation of the molecular motion in liquids.[26]

## VI. CONCLUSIONS

We have revisited the GR microscopic theory of molecular motion in classical monatomic liquids where an equation of motion for the VAF is derived by assuming a Brownian particle diffusing in a mean-time-dependent harmonic force field. The original GR theory has been extended using a non-Markovian Langevin equation incorporating the dynamic friction. We have demonstrated that the inclusion of dynamic friction not only overcomes the limitations of the GR theory but, also provides self-consistent information about the time-dependence of the friction coefficient. It has been shown that the dynamic friction decays exponentially at low and moderate liquid densities whereas it exhibits exponential growth in high density liquids like liquid metals. Amidst the growing interest and efforts to understand the molecular origin of the dynamic friction and; its implications on microscopic transport properties [7, 52, 56], our findings about the dynamic friction in liquids provides a new outlook to the Brownian description of atomic dynamics in liquids. The modified equation of motion for the VAF (Eq. 21), that includes the dynamic friction, gives excellent account of the velocity correlations in a broad range of liquid densities where the physical scenario changes from binary collision dominated short-time dynamics to the one with significant many-body cooperative effects. The results for high density liquids, especially the observed negative molecular relaxation rates ($\alpha$), intuitively hints at the existence of imaginary eigenmodes (unstable modes) in the density of states of the INMs in accordance with Zwanzig's disordered-solid-like picture of short-time dynamics. However, an elaborate quantitative analysis and physical interpretation of this aspect is constrained due to non-availability of a tangible analytical solution of Eq. (21). This is the primary issue that yet remains to be addressed to acquire in-depth understanding of the time-dependence of the dynamical friction and its implications on the dynamical correlations at short-times in liquids.

**ACKNOWLEDGEMENTS**

KNL gratefully acknowledge the financial support from SERB, New Delhi for the research project CRG/2018/001552 (2019-2022). We are grateful to Dr. J. N. Pandya, Department of Applied Physics, Faculty of Technology and Engineering, The M. S. University of Baroda, Vadodara, for his guidance and help in obtaining numerical solutions using Mathematica.

# APPENDIX

# MOLECULAR DYNAMICS SIMULATION DETAILS

Classical molecular simulations on Cu, Fe and Ni have been performed using the LAMMPS code [57] employing Finnis-Sinclair type many-body potential for Cu [58] and embedded-atom model (EAM) potentials for Ni [59] and Fe [60]. All the investigated metals were subjected to the same simulation protocol described here. Total 4000 atoms were taken in a cubic simulation box subject to the periodic boundary conditions. The equations of motion were solved using Verlet's algorithm in the velocity form with a timestep of 2 fs in NPT ensemble. An equilibrated liquid configuration for the metal at 2000 K was obtained by melting it and homogenizing it for 0.6 ns. The liquid metal at 2000 K was subsequently quenched to 300 K at a rate of 0.1 K/ps under zero pressure condition in NPT ensemble. The liquid configurations at intermediate temperatures of interest were recorded during the quench run and, further equilibrated thoroughly to extract the structural and dynamical information. Separate production runs at different temperatures were performed to record atomic trajectories to obtain the pair correlation function and the velocity autocorrelation function.

**Table Caption**

**Table I:** Parameters for LJ system (T = 1.5). All dimensionless quantities should be multiplied by an appropriate combination of MD units (σ, ε and the particle mass, *m*). Conversion to argon values can be more appropriately done using $\frac{\varepsilon}{k_B} = 155.876\ K$ and $\sigma = 3.40\ Å$.

**Table II:** Parameters for liquid alkali metals. $\langle \nabla^2 V \rangle$ has been derived using Eq. (7). The derivatives of effective pair potentials and the pair correlation function is discussed in the text.

**Table III:** Parameters of liquid transition metals. $\langle \nabla^2 V \rangle$ has been derived using Eq. (7). The derivatives of effective pair potentials and the pair correlation function is discussed in the text.

**Figure Caption**

Figure 1: VAF for LJ systems at T = 1.5 and different densities (a) 0.20, (b) 0.40 and, (c) 0.84. Inset in (c) demonstrates that the results for α > 0 are way off the MD results. MD results [Ref 36] (d) Residue plot showing comparison of the quality of results obtained using different models for ρ = 0.84. (e) Residue plot depicting the deviations of the results of present model from MD results at different densities.

Figure 2: VAF for liquid lithium above melting point. (a) 725 K, (b) 574 K and, (c) 470 K. MD results [41]

Figure 3: VAF for liquid Na. (a) 900 K, (b) 380 K MD results [Ref. 42]

Figure 4: VAF for liquid potassium at 450 K. MD results [Ref. 43]

Figure 5: VAF for liquid Cu. (a) 1873 K, (b) 1773 K, (c) 1573 K, (d) 1423 K. MD results (Present work) Details MD simulations are given in Appendix

Figure 6: VAF for liquid Ni. (a) 1923 K, (b) 1810 K, (c) 1735 K, (d) 1500 K. MD results (Present work) Details MD simulations are given in Appendix

Figure 7: VAF for liquid Fe. (a) 2023 K, (b) 1923 K, (c) 1873 K, (d) 1833 K. MD results (Present work) Details MD simulations are given in Appendix

**Table I**

| $\rho$ | $(\omega_0)_{fit}$ ($10^{12}$ s$^{-1}$) | $(\beta_0)_{fit}$ ($10^{12}$ s$^{-1}$) | $\alpha$ ($10^{12}$ s$^{-1}$) |
|---|---|---|---|
| 0.2 | 0.55 ± 0.03 | 3.69 ± 0.14 | 2.03 ± 0.03 |
| 0.4 | 7.76 ± 0.03 | 28.9 ± 0.27 | 3.69 ± 0.02 |
| 0.84 | 11.71 ± 0.04 | 14.79 ± 0.09 | -1.17 ± 0.06 |

**Table II**

| System | T (K) | $\langle \nabla^2 V \rangle$ ($10^2$ J m$^{-2}$) | $\omega_0$ ($10^{12}$ s$^{-1}$) | $\beta_0$ ($10^{12}$ s$^{-1}$) | $(\omega_0)_{fit}$ ($10^{12}$ s$^{-1}$) | $(\beta_0)_{fit}$ ($10^{12}$ s$^{-1}$) | $\alpha$ ($10^{12}$ s$^{-1}$) |
|---|---|---|---|---|---|---|---|
| Li | 470 | 0.7726 | 47.27 | 27.4 | 42.0 | 30.5 | -6.0 |
| | 574 | 0.7452 | 46.42 | 61.92 | 49.0 | 42.0 | -2.5 |
| | 725 | 0.7351 | 46.75 | 44.75 | 49.0 | 42.0 | -2.5 |
| Na | 380 | 0.2049 | 13.38 | 7.67 | 16.5 | 11.0 | -3.2 |
| | 900 | 0.3646 | 17.84 | 25.60 | 18.9 | 23.8 | -2.5 |
| K | 450 | 0.2054 | 10.27 | 6.73 | 9.8 | 6.4 | -3.9 |

**Table III**

| System | T (K) | $\langle \nabla^2 V \rangle$ ($10^2$ J m$^{-2}$) | $\omega_0$ ($10^{12}$ s$^{-1}$) | $\beta_0$ ($10^{12}$ s$^{-1}$) | $(\omega_0)_{fit}$ ($10^{12}$ s$^{-1}$) | $(\beta_0)_{fit}$ ($10^{12}$ s$^{-1}$) | $\alpha$ ($10^{12}$ s$^{-1}$) |
|---|---|---|---|---|---|---|---|
| Cu | 1423 | 2.2756 | 26.8117 | 12.9671 | 25.0 | 15.0 | -6.7 |
| | 1573 | 2.2762 | 26.8152 | 15.0213 | 27.5 | 19.0 | -4.5 |
| | 1773 | 2.2757 | 26.8123 | 20.8056 | 25.0 | 18.0 | -6.5 |
| | 1873 | 2.2787 | 26.7710 | 22.2103 | 25.0 | 19.0 | -6.5 |
| Ni | 1500 | 3.2057 | 21.9920 | 11.5358 | 34.0 | 21.5 | -8.7 |
| | 1735 | 3.1137 | 21.5839 | 14.6925 | 34.0 | 24.0 | -6.7 |
| | 1810 | 3.0794 | 21.4368 | 18.7751 | 34.0 | 25.0 | -6.7 |
| | 1923 | 3.0223 | 21.1881 | 18.8385 | 34.0 | 25.0 | -6.7 |
| Fe | 1833 | 3.2529 | 34.1950 | 15.7441 | 34.0 | 23.0 | -6.7 |
| | 1873 | 3.2320 | 34.0850 | 15.8361 | 34.0 | 24.0 | -6.9 |
| | 1923 | 3.2182 | 34.0122 | 16.2939 | 34.0 | 24.0 | -7.9 |
| | 2023 | 3.1823 | 33.8219 | 19.1028 | 34.0 | 24.0 | -8.9 |

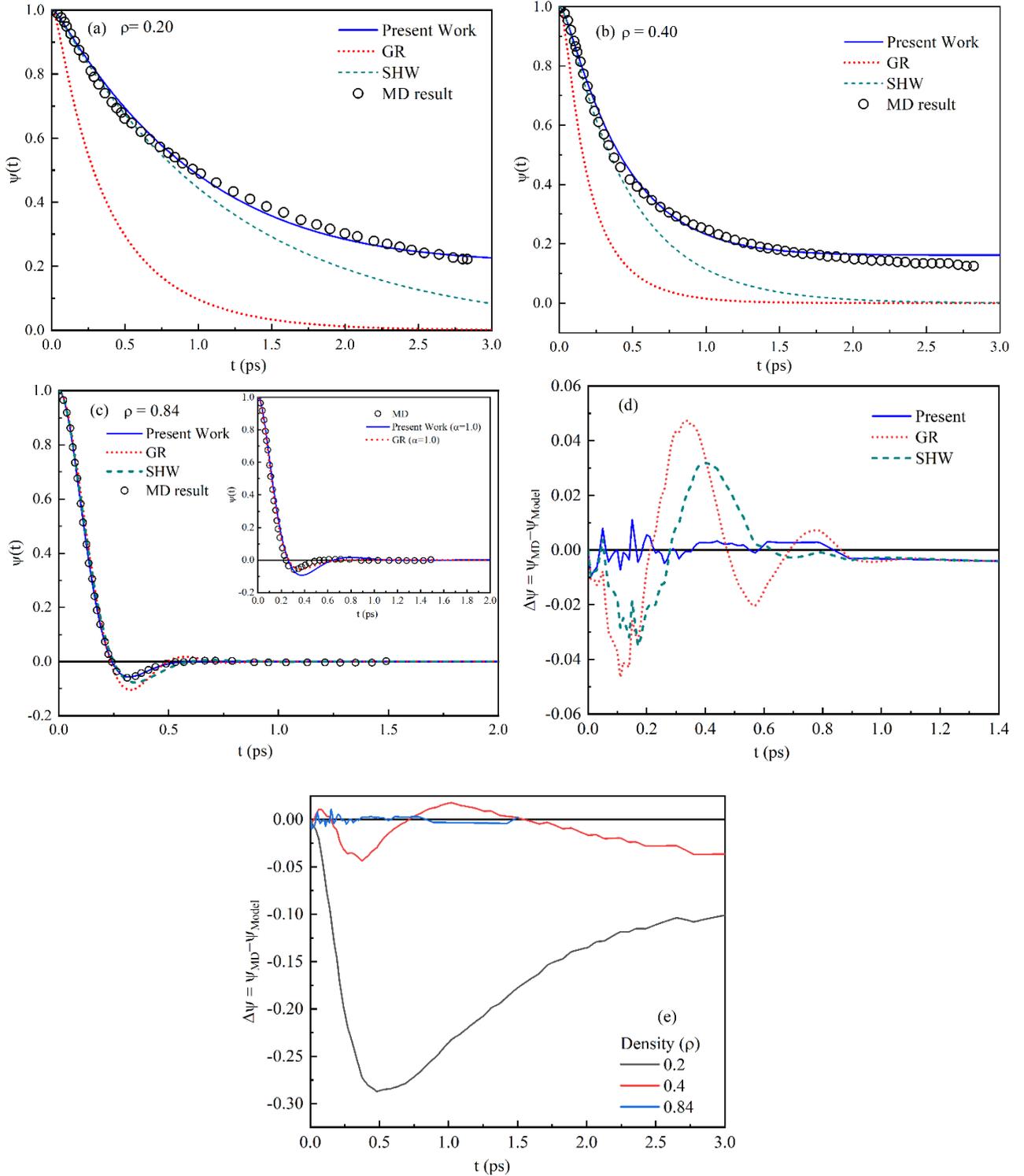

Fig. 1: VAF for LJ systems at T = 1.5 and different densities (a) 0.20, (b) 0.40 and, (c) 0.84. Inset in (c) demonstrates that the results for α > 0 are way off the MD results. MD results [Ref 36] (d) Residue plot showing comparison of the quality of results obtained using different models for ρ = 0.84. (e) Residue plot depicting the deviations of the results of present model from MD results at different densities.

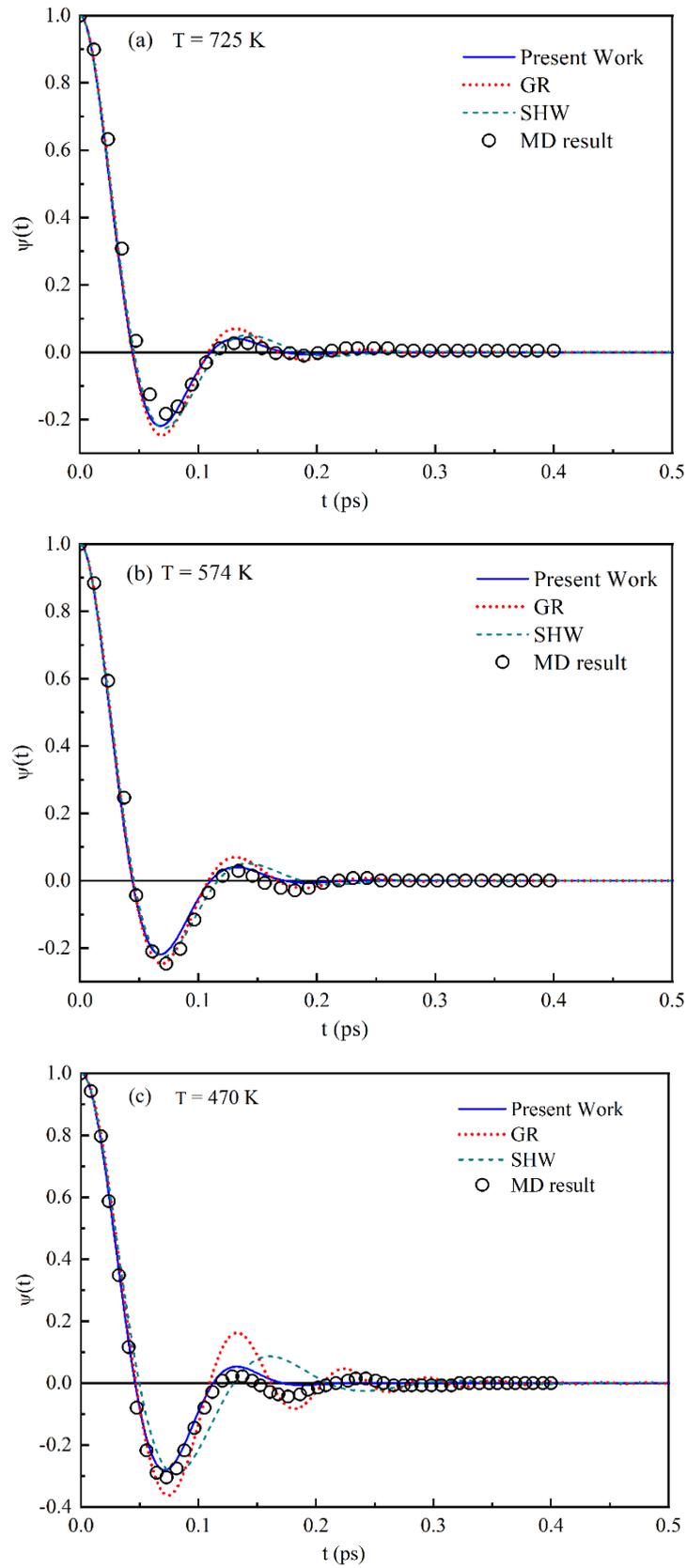

Fig. 2: VAF for liquid lithium above melting point. (a) 725 K, (b) 574 K and, (c) 470 K. MD results [Ref. 41]

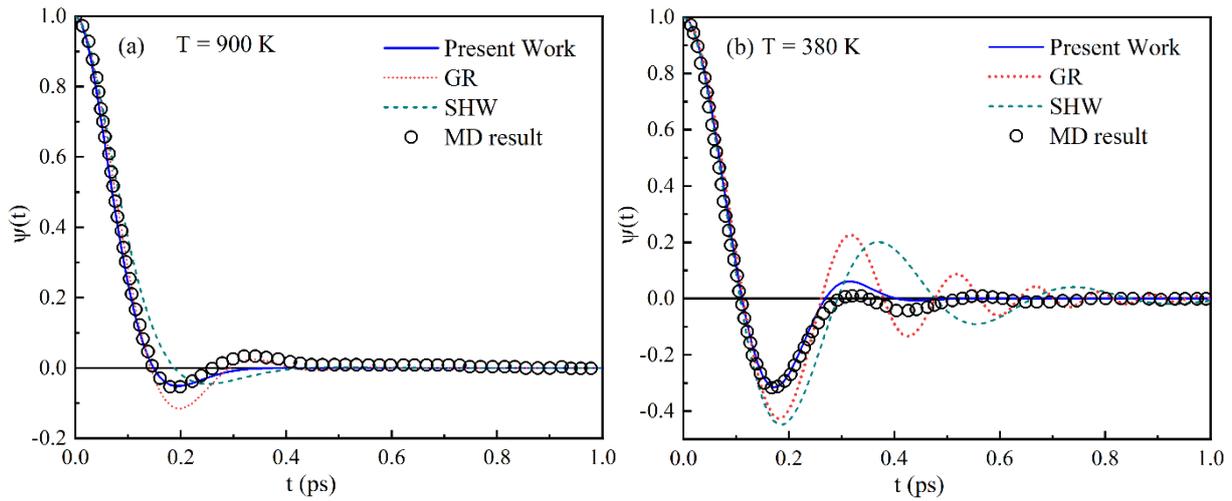

Fig. 3: VAF for liquid Na. (a) 900 K, (b) 380 K MD results [Ref. 42]

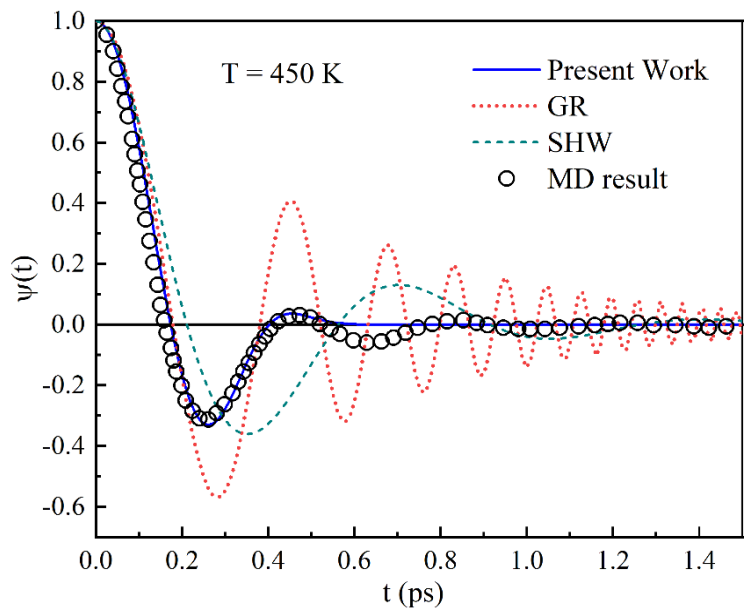

Fig. 4: VAF for liquid potassium at 450 K. MD results [Ref. 43]

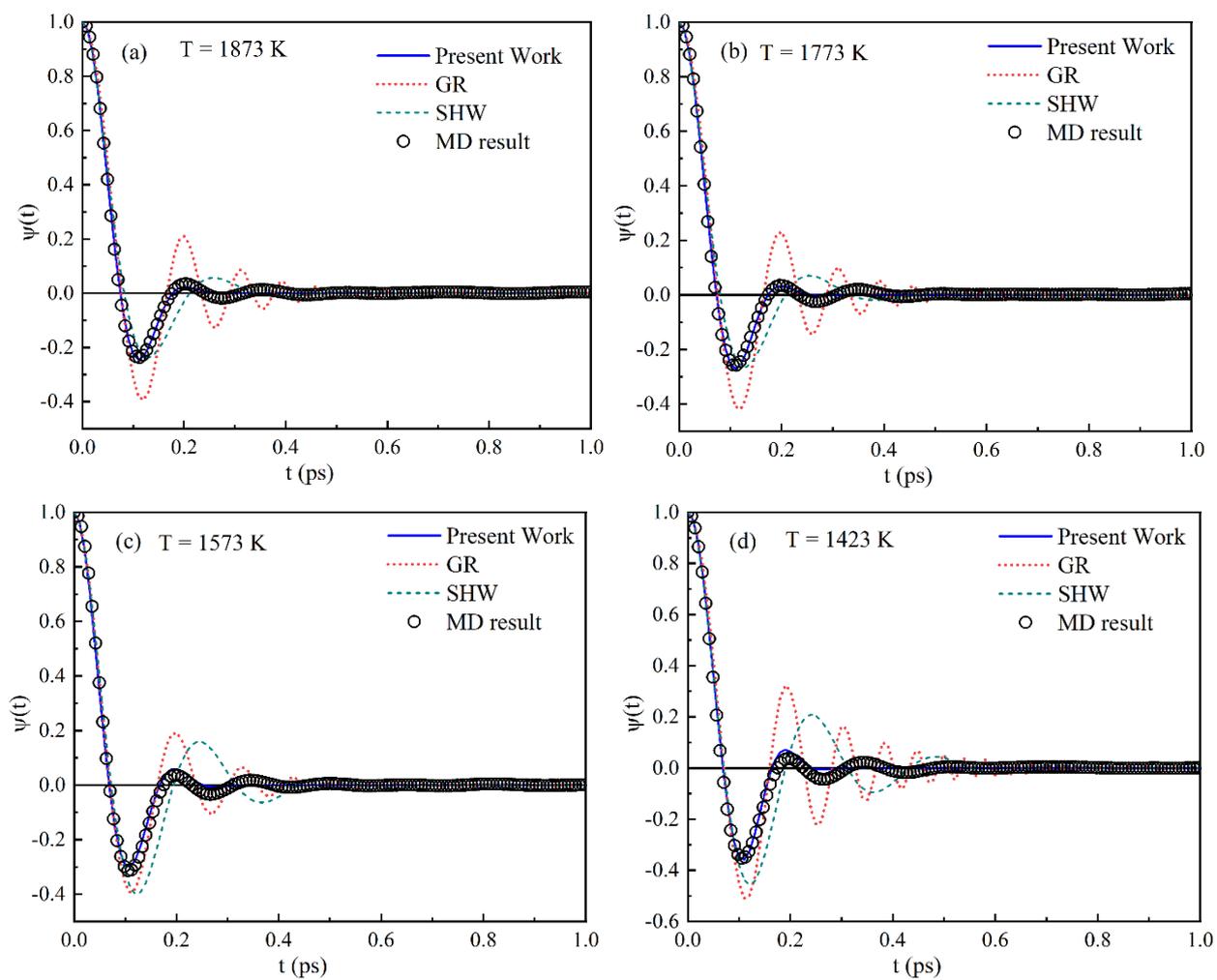

Fig. 5: VAF for liquid Cu. (a) 1873 K, (b) 1773 K, (c) 1573 K, (d) 1423 K. MD results (Present work)

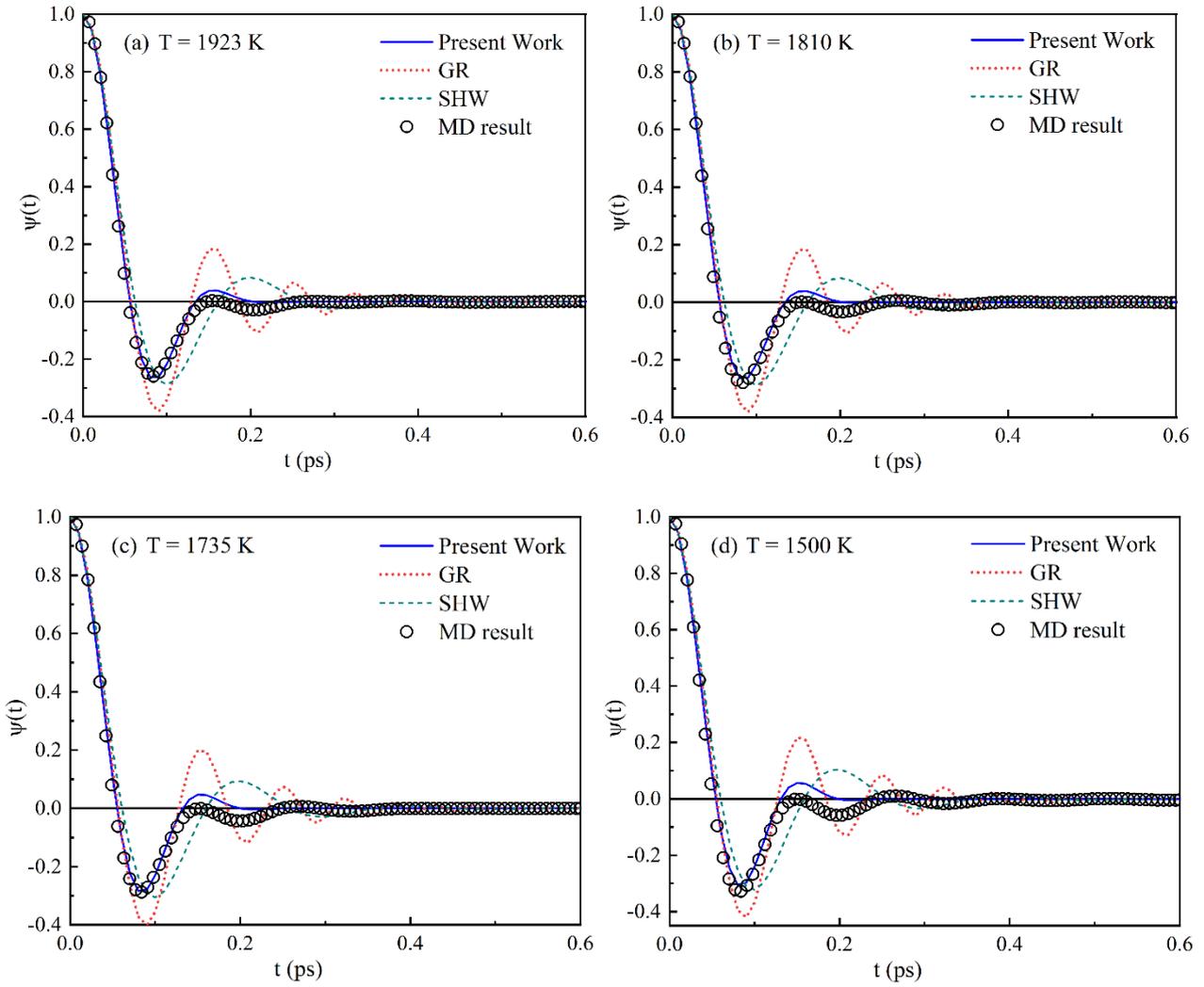

Fig. 6: VAF for liquid Ni. (a) 1923 K, (b) 1810 K, (c) 1735 K, (d) 1500 K. MD results (Present work)

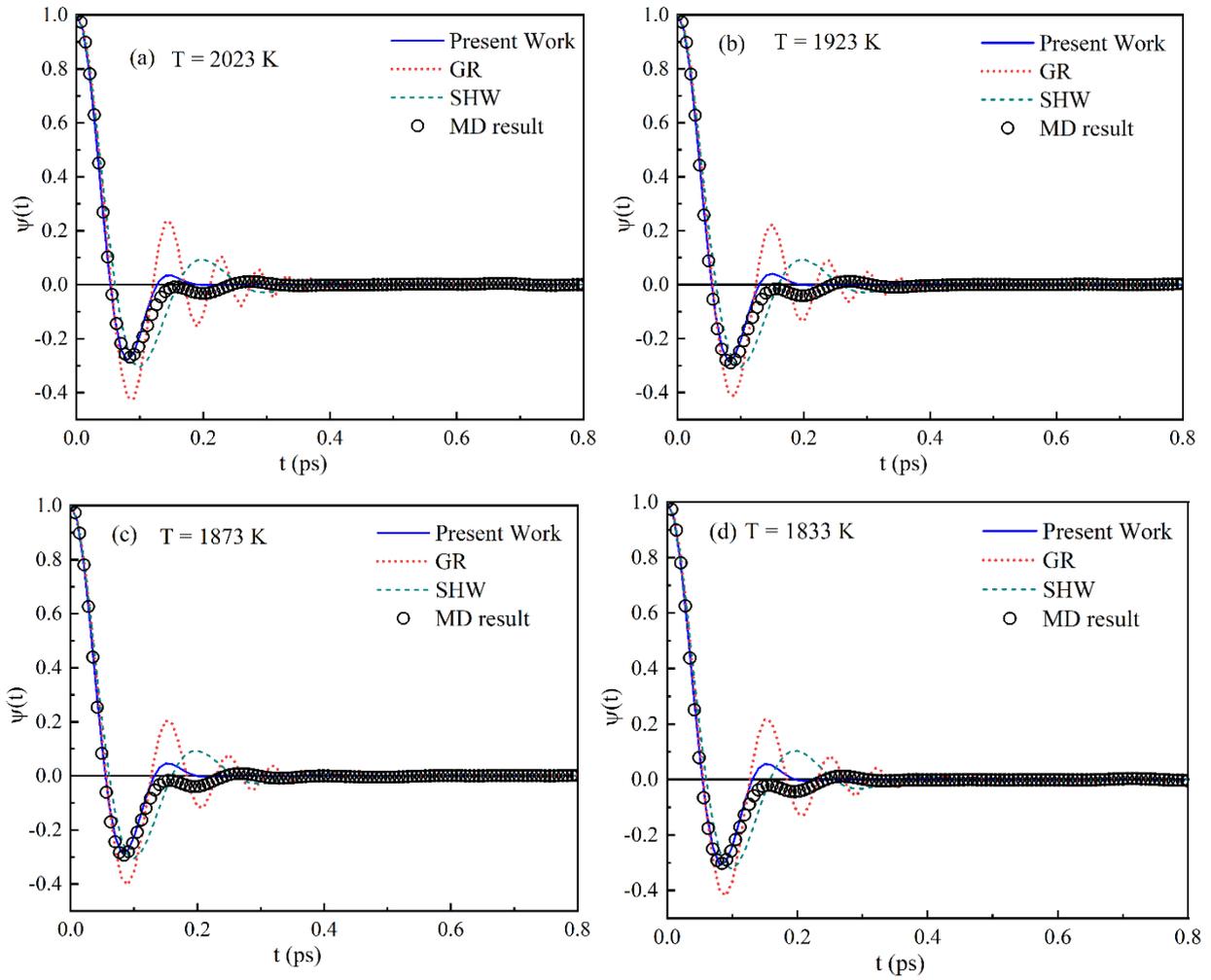

Fig. 7: VAF for liquid Fe. (a) 2023 K, (b) 1923 K, (c) 1873 K, (d) 1833 K. MD results (Present work)